\documentclass[12pt]{iopart}
\usepackage{graphicx}
\usepackage{xspace}
\usepackage[breaklinks]{hyperref}
\usepackage{color}
\newcommand{\srsl}{Sr$_2$IrO$_4$\xspace}

\begin{document}
\rapid[Locking of iridium magnetic moments to the correlated rotation of oxygen octahedra in \srsl]{Locking of iridium magnetic moments to the correlated rotation of oxygen octahedra in \srsl revealed by X-ray resonant scattering}

\author{S Boseggia$^1$$^,$$^2$, H C Walker$^3$, J Vale$^1$$^,$$^4$, R Springell$^5$, Z Feng$^1$, R S Perry$^6$, M. Moretti Sala$^7$, H.~M. R\o{}nnow$^4$, S P Collins$^2$ and D F McMorrow$^{1,8}$}
\address{$^1$ London Centre for Nanotechnology and Department of Physics and Astronomy, University College London, London WC1E 6BT, UK}
\address{$^2$ Diamond Light Source Ltd, Diamond House, Harwell Science and Innovation Campus, Didcot, Oxfordshire OX11 0DE, UK}
\address{$^3$ Deutsches Elektronen-Synchrotron DESY, 22607 Hamburg, Germany}
\address{$^4$ Laboratory for Quantum Magnetism, ICMP, \'Ecole Polytechnique F\'ed\'erale de Lausanne (EPFL), CH-1015 Lausanne, Switzerland}
\address{$^5$ Royal Commission for the Exhibition of 1851 Research Fellow, Interface Analysis Centre, University of Bristol BS2 8BS,UK}
\address{$^6$ Scottish Universities Physics Alliance, School of Physics, University of Edinburgh, Mayfield Road, Edinburgh EH9 3JZ, Scotland}
\address{$^7$ European Synchrotron Radiation Facility, BP 220, F-38043 Grenoble Cedex, France}
\address{$^8$ Department of Physics, Technical University of Denmark, DK-2800 Kgs. Lyngby, Denmark}
\ead{stefano.boseggia@diamond.ac.uk}
\begin{abstract}
\srsl is a prototype of the class of Mott insulators in the strong spin-orbit interaction (SOI) limit described by a $J_{\mathrm{eff}}=1/2$ ground state. In \srsl, the strong SOI is predicted to manifest itself in the locking of the canting of the magnetic moments to the correlated rotation by 11.8(1)$^{\circ}$ of the oxygen octahedra that characterizes its distorted layered perovskite structure. Using X-ray resonant scattering at the Ir $L_3$ edge we have measured accurately the intensities of Bragg peaks arising from different components of the magnetic structure. From a careful comparison of integrated intensities of peaks due to  basal-plane antiferromagnetism, with those due to $b$-axis ferromagnetism, we deduce a canting of the magnetic moments of 12.2(8)$^{\circ}$. We thus confirm that in \srsl the magnetic moments rigidly follow the rotation of the oxygen octahedra, indicating that, even in the presence of significant non-cubic structural distortions, it is a close realization of the $J_{\mathrm{eff}}=1/2$ state.

\end{abstract}

\submitto{\JPC}
\maketitle
Recently, Ir-based transition metal oxides have been identified as a fertile ground  to search for novel correlated ground states and excitations \cite{Pesin,axion_PhysRevB.83.205101,PhysRevLett.105.027204}. The salient interactions in these compounds ---  the electronic band width, on-site Coulomb repulsion, and spin-orbit --- are on an similar energy scale. Layered iridate perovskites, in particular, have attracted significant interest for their structural and magnetic similarities to layered cuprates, and for the first observation of a spin-orbit induced $J_{\mathrm{eff}}=1/2$ Mott-like insulating state in \srsl  \cite{Kim-PhysRevLett-2008}. In this scenario, the spin-orbit interaction (SOI) plays a fundamental role by narrowing the 5$d$ band width, permitting a modest (compared to 3$d$ ions) on-site Coulomb repulsion to produce a Mott insulating ground state. Due to the entanglement between the spin and orbital momenta, the $J_{\mathrm{eff}}=1/2$ state shows an intriguing three dimensional shape \cite{PhysRevLett.102.017205} that produces unusual magnetic interactions as a function of the dimensionality \cite{PhysRevLett.109.037204,PhysRevB.85.184432,0953-8984-24-31-312202,PhysRevLett.109.157402},and the variation of the local symmetry \cite{PhysRevLett.110.117207}.

\begin{figure}
\centering
\includegraphics[width=1\textwidth]{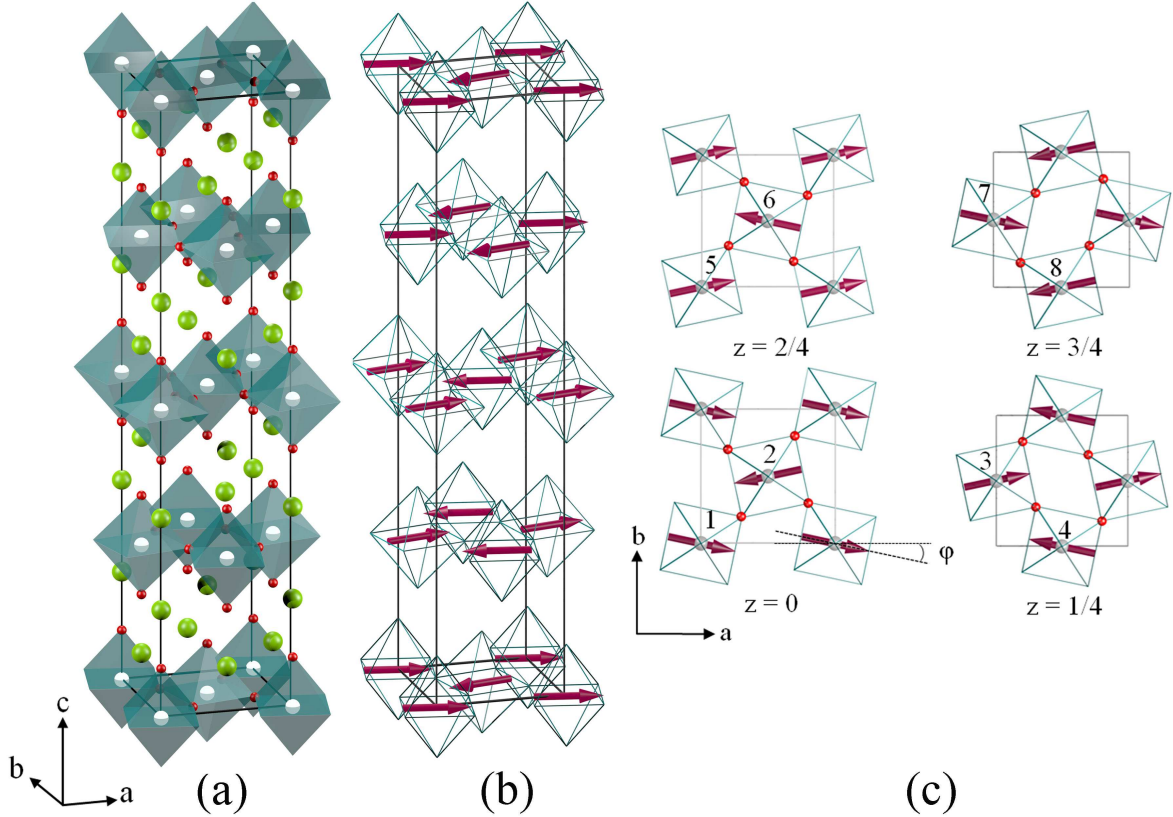}
\caption{(Color Online) (a) Crystal structure of \srsl (space group I4$_1$/\textit{acd}). IrO$_6$ layers where the Ir atoms (grey) are at the center of corner sharing oxygen (red) octahedra are separated by Sr atoms (light green). The IrO$_6$ octahedra undergo a staggered correlated rotation of $\sim 12^{\circ}$ about the $c$ axis. (b) \textit{ab} plane canted magnetic structure. The magnetic moment canting angle follows  the octahedral rotations rigidly. (c) Magnetic stacking pattern along the $c$ axis. Sites 1-8 in the space group I4$_1$/\textit{acd} (\#142, origin choice 1) for the Ir ions were used to calculate the XRMS cross section. Ir magnetic moments are canted by an angle $\phi$ from the $a$-axis}
\label{fig:MagStruct}
\end{figure}

The magnetic structure of \srsl has been extensively investigated in a number of experimental and theoretical studies \cite{Kim-Science-2009,PhysRevLett.102.017205,PhysRevB.87.144405,PhysRevLett.110.117207,0953-8984-23-25-252201}. On the basis of
their pioneering X-ray resonant magnetic scattering (XRMS) experiments, Kim \textit{et al.}  \cite{Kim-Science-2009} assigned a $J_{\mathrm{eff}}=1/2$ groundstate to \srsl, in support of a previous  proposal \cite{Kim-PhysRevLett-2008}. However, this interpretation has been questioned \cite{0953-8984-23-25-252201,2013arXiv1308.0128M,PhysRevLett.109.027204}, clearly calling for further elucidation of the true nature of the ground state, such as might be achieved by a complete understanding of its magnetic structure. Although Kim \textit{et al.} did not undertake a precise study of the magnetic structure of \srsl, they did establish some of its essential features (see Figure \ref{fig:MagStruct}): the moments lie in the basal plane and form a canted  antiferromagnetic (AF) structure. We later refined this picture by employing azimuthal scans of magnetic Bragg peaks to show that the AF component of the order is along the $a$-axis in the I4$_1$/\textit{acd} reference system \cite{PhysRevLett.110.117207}. Neither of these XRMS studies addressed the issue of the magnitude of the canting angle $\phi$, although a recent neutron diffraction investigation has addressed the point, reporting a canting angle $\phi=13(1)^{\circ}$   \cite{PhysRevB.87.140406}.  From the theoretical point of view, the experimental determination of the canting angle places a strong constraint on the Hamiltonian and the theoretical model of perovskite iridates. An effective Hamiltonian including the both tetragonal crystal field ($\Delta$) and octahedral rotation ($\rho$) has been derived by Jackeli and Khaliullin \cite{PhysRevLett.102.017205}. According to this model, in the strong SOI limit (for $\Delta\rightarrow 0$), the ratio of the magnetic moment canting angle to the IrO$_6$ octahedral rotation ($\phi/\rho$) approaches unity.

Here we report measurements of the magnetic structure of \srsl which have been designed to accurately determine the canting angle of the magnetic moments. We find that the Ir magnetic moments deviate by 12.2(8)$^{\circ}$ from the $a$-axis, following the octahedral rotation of about 11.8$^{\circ}$, rigidly. Our study supports the full realization of a $J_{\mathrm{eff}}=1/2$ state in \srsl, and the marginal effect of the small tetragonal distortion on its magnetic properties.

\begin{figure}
\centering
\includegraphics[width=0.5\textwidth]{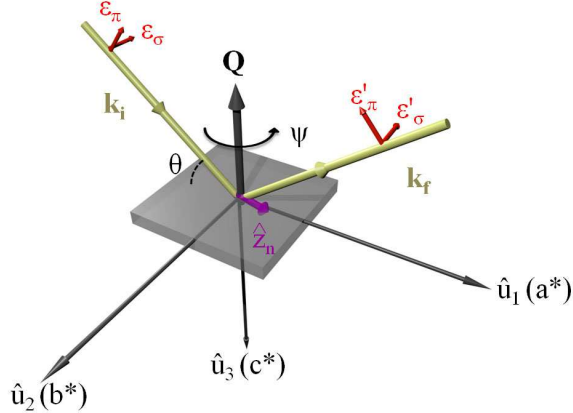}
\caption{ (Color Online) Scattering geometry used in the present experiment. The
sample has been oriented with the [0\,0\,1] and the [1\,0\,0] directions lying in the scattering plane, defined by the incoming and
outgoing wave vectors. A vertical scattering geometry with a $\sigma$ polarized incident beam was exploited. $\theta$ is the Bragg angle, $\psi$ represents an azimuthal rotation about the scattering vector $\mathbf{Q}$.}
\label{fig:ScattGeometry}
\end{figure}

The XRMS experiment was conducted at the I16 beamline at Diamond Light Source, Didcot, UK. A monochromatic x-ray beam at the Ir $L_3$ edge (11.217 keV) was provided by means of a U27 undulator insertion device and a channel-cut Si (1\,1\,1) monochromator. In order to detect the scattered photons an avalanche photodiode (APD) was exploited, together with a Au (3\,3\,3) crystal to analyze the polarization of the scattered beam. The \srsl sample was mounted in a closed-circle cryostat with the [0\,0\,1] (perpendicular to the sample surface) and [1\,0\,0] directions in the vertical scattering plane of a Newport 6-circle  Kappa diffractometer at the azimuthal origin. The scattering geometry is illustrated in Figure \ref{fig:ScattGeometry}.

Single crystals of \srsl were grown as in reference  \cite{PhysRevLett.110.117207}. \srsl crystallizes in tetragonal space group I4$_1$/\textit{acd} ($a=b=5.48 ~\AA$, and $c=25.8 ~\AA$)  \cite{PhysRevB.49.9198}. Although the exact space group was recently called into question by two different neutron studies  \cite{PhysRevB.87.140406,PhysRevB.87.144405}, we note that the subtle difference from the commonly used I4$_1$/\textit{acd} is not relevant in terms of the magnetic structure and therefore the I4$_1$/\textit{acd} reference will be used in our study. The peculiarity of this crystal structure is that the Ir$^{4+}$ ions lie at the centre of slightly elongated  (4\%) octahedra  that are alternately rotated by 11.8$^{\circ}$ about the $c$-axis (see Figure \ref{fig:MagStruct}(a))  \cite{PhysRevB.49.9198}. From the theoretical point of view, a strong link between the crystal and magnetic structure is predicted. Due to strong spin-orbit coupling and the tilting of the IrO$_6$ octahedra, a Dzyaloshinsky-Moriya (DM) interaction arises. However, the anisotropy of the single-layer compound can be gauged away by proper site-dependent spin rotations. The twisted Hubbard model can then be mapped onto a SU(2)-invariant pseudospin-1/2 system, being isostructural with, for instance, the close relative Ba$_2$IrO$_4$. Here, the straight Ir-O-Ir bonds preserve inversion symmetry so that the system shows a simple basal-plane antiferromagnetic structure\cite{PhysRevLett.110.117207}. To obtain the magnetic structure of the twisted system, we have to transform the isotropic system back. As a result, the spins are canted exactly like the IrO$_6$ octahedra.
The magnetic structure of \srsl, illustrated in Figure \ref{fig:MagStruct}(b-c), can be  decomposed  into a basal-plane antiferromagnetic sublattice A, where the moments are pointing along the [1\,0\,0] direction, and  a net $b$-axis ferromagnetic moment due to canting of Ir magnetic moments, that generates a stacked antiferromagnetic structure $(-++-)$ along the $c$-axis (sublattice B). The A sublattice is responsible for the ($1$\,$0$\,$4n$) and ($0$\,$1$\,$4n\!+\!2$) magnetic peaks, and the B for the (0\,0\,$2n\!+\!1$) magnetic peaks. The relative intensity of the magnetic reflections associated with the two magnetic sublattices  is ultimately linked to the projection of the magnetic moment on the $a$ and $b$ axes, respectively. In the total absence of the $b$-axis ferromagnetic component, the intensity of the (0\,0\,$2n\!+\!1$) magnetic reflection vanishes, as in the Ba iridate case  \cite{PhysRevLett.110.117207}. It is therefore possible to infer the direction of Ir magnetic moments from the comparison between a theoretical model for the magnetic moments and the measured intensity ratio $I_A/I_B$ between several magnetic reflections.

\begin{figure}
\centering
\includegraphics[width=0.5\textwidth]{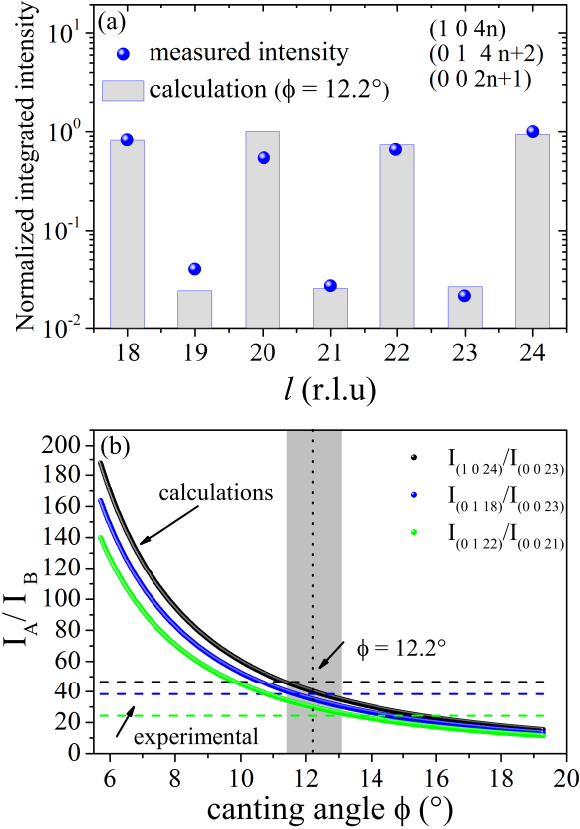}
\caption{(Color Online) (a) $l$-scan across the (1\,0\,$4n$), (0\,1\,$4n\!+\!2$), and (0\,0\,$2n\!+\!1$) reflections at $T=10\,\mathrm{K}$ at the Ir $L_3$ edge. The blue spheres represent the integrated intensity of the measured magnetic scattering corrected for absorption as discussed in the text. The heights of the light blue bars represent the calculated intensity for the magnetic moment arrangement of Figure \ref{fig:MagStruct}(c) for a canting angle $\phi=12.2^{\circ}$. (b) Intensity ratio ($I_{\mathrm{A}}/I_{\mathrm{B}}$) between the $a$-axis in-plane antiferromagnetic reflections (magnetic sublattice A) and the $b$-axis canting-induced magnetic reflections (magnetic sublattice B) calculated as a function of the canting angle $\phi$ (solid points). The dashed lines are the experimental intensity ratios  at the Ir $L_3$ edge.}
\label{fig:CantingAngle}
\end{figure}
\begin{figure}
\centering
\includegraphics[width=0.5\textwidth]{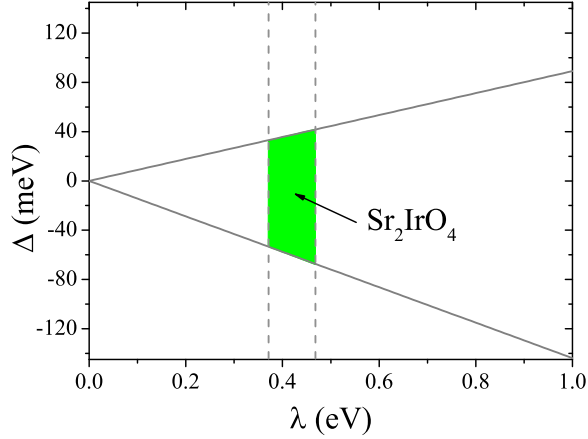}
\caption{(Color Online) Tetragonal crystal field parameter ($\Delta$) as a function of the spin-orbit coupling $\lambda$. The boundaries of the tetragonal crystal field were obtained from the Hamiltonian of Reference\cite{PhysRevLett.102.017205} using, as an input parameter, the error bar in the canting angle $\phi$ obtained from the present study.}
\label{fig:CF}
\end{figure}
With the energy of the incoming photons tuned at the Ir $L_3$ edge we measured several magnetic peaks along the ($1$\,$0$\,$4n$), ($0$\,$1$\,$4n\!+\!2$), and (0\,0\,$2n\!+\!1$) directions in the $\sigma$-$\pi$ polarization channel at $T=10$ K. $\theta$-$2\theta$ scans of each magnetic peak were numerically integrated and corrected for self-absorption by multiplying the observed intensity by the factor:
\begin{equation}
Abs(\mathbf{Q},\psi)=1+\frac{\sin \alpha(\mathbf{Q},\psi)}{\sin \beta(\mathbf{Q},\psi)},
\end{equation}
where $\alpha(\mathbf{Q},\psi)$ and $\beta(\mathbf{Q},\psi)$ are respectively the incident and exit angle with respect to the (0\,0\,1) sample surface. The results as a function of $l$ are plotted in Figure \ref{fig:CantingAngle}(a) (blue spheres).

In order to interpret the experimental data, and to extract the canting angle of Ir magnetic moments, we calculated the resonant scattering cross section for the magnetic moment arrangement of Figure \ref{fig:MagStruct}(c). Following the formalism developed by Blume and Gibbs \cite{PhysRevB.37.1779}, and Hill and McMorrow  \cite{Hill:sp0084} we can write the E1-E1 resonant magnetic scattering amplitude  as:
\begin{equation}
f_{n\mathrm{E1}}^{\mathrm{XRMS}}=-\imath F^{(1)}_{\mathrm{E1}}(\mathbf{\hat{\epsilon}} '\times \mathbf{\hat{\epsilon}})\cdot\mathbf{\hat{z}_n}
\end{equation}
where $\mathbf{\hat{\epsilon}}$ ($\mathbf{\hat{\epsilon}}'$) is the incoming (scattered) x-ray linear polarization orientation, and $\mathbf{\hat{z}_n}$ is a unit vector in the direction of the Ir magnetic moments (see Figure \ref{fig:ScattGeometry}). $F^{(1)}_{\mathrm{E1}}$ are coefficients dependent on the electronic transitions that determine the strength of the resonant process  \cite{PhysRevLett.61.1245}.

We can then calculate the magnetic scattering amplitude for the Ir magnetic moments pointing along the $a$-axis (sublattice A) as:

\begin{equation}
f_{n\mathrm{E1}_{\mathrm{A}, (1 0 4n)}}^{\mathrm{XRMS}}=\imath \mathbf{\hat{z}_n}\cos\phi\left( \cos\xi\cos\theta\cos\psi +\sin\theta \sin\xi\right)
\end{equation}
and
\begin{equation}
f_{n\mathrm{E1}_{\mathrm{A}, (0 1 4n+2)}}^{\mathrm{XRMS}}=\imath \mathbf{\hat{z}_n}\cos\phi\cos\psi\cos\theta,
\end{equation}
and the magnetic scattering amplitude for the Ir magnetic moments pointing along the $b$-axis as
\begin{equation}
f_{n\mathrm{E1_{B}}}^{\mathrm{XRMS}}=\imath\mathbf{\hat{z}_n} \sin\phi \sin \psi \cos\theta,
\end{equation}
where $\phi$ is the canting angle as defined in Figure \ref{fig:MagStruct}(c), $\psi$ is the azimuthal rotation about the scattering vector $\mathbf{Q}$, $\theta$ is the Bragg angle, and $\xi$ is the angle between the scattering vector $\mathbf{Q}$ and the $c$-axis.
The total scattering cross section is then calculated as the squared modulus of the amplitude, taking into account the phase factor deriving from the magnetic structure factor as
\begin{equation}
I\propto\left| \sum_n e^{2\pi \imath\mathbf{Q}\cdot\mathbf{r}_n}f_{n\mathrm{E1}}^{\mathrm{XRMS}}\right|^2,
\end{equation}
where $\mathbf{Q}$ is the scattering vector and  $\mathbf{r}_n$ is the crystallographic coordinate of the $n$th Ir ion. The positions of the eight Ir ions over which the sum runs  are illustrated in Figure \ref{fig:MagStruct}(c).
We can now calculate the resonant cross section for the reflections of the two magnetic sublattices as:
\begin{equation}
I_{\mathrm{A}, (1 0 4n)}\propto\left| + 2\left(1+e^{\frac{\imath\pi l}{2}}\right)\left(1+e^{\imath\pi l}\right)\imath \mathbf{\hat{z}_n}\cos\phi\left( \cos\xi\cos\theta\cos\psi +\sin\theta \sin\xi\right)\right|^2,
\end{equation}
\begin{equation}
I_{\mathrm{A}, (0 1 4n+2)}\propto\left| - 2\left(-1+e^{\frac{\imath\pi l}{2}}\right)\left(1+e^{\imath\pi l}\right)\imath \mathbf{\hat{z}_n}\cos\phi\cos\psi\cos\theta\right|^2,
\end{equation}
and
\begin{equation}
I_{\mathrm{B}}\propto\left| -2\left(-1+ e^{\frac{\imath\pi l}{2}}\right)^2\left(1+e^{\frac{\imath\pi l}{2}}\right)\imath\mathbf{\hat{z}_n}\sin\phi \sin \psi \cos\theta\right|^2.
\end{equation}

Figure \ref{fig:CantingAngle}(b) shows the comparison between the calculated intensity ratio $I_A/I_B(\phi)$ and the experimental value for five different magnetic reflections: (1\,0\,24), (0\,1\,22), (0\,1\,18) associated with sublattice A, and (0\,0\,21) and (0\,0\,23) associated with the canting-induced magnetic sublattice B. From the intersection between the calculated $I_A/I_B(\phi)$ curves and the observed value we can deduce the canting angle of the  Ir magnetic moments. A deviation from the $a$-axis by 12.2(8)$^{\circ}$ is obtained averaging  the canting angle associated with the three intensity ratios $I_{(1\,0\,24)}$/$I_{(0\,0\,23)}$, $I_{(0\,1\,18)}$/$I_{(0\,0\,23)}$, and $I_{(0\,1\,22)}$/$I_{(0\,0\,21)}$ (see Figure \ref{fig:CantingAngle}(b)). We therefore conclude that, within the experimental error, the  magnetic moments in \srsl follow  the octahedral rotations rigidly. We note that from our analysis we cannot determine the sign of the canting angle $\phi$. Based on the prediction of theoretical models for iridate perovskites, we exclude that the  Ir moments could rotate in antiphase with the oxygen octahedra.
Our findings support the Hamiltonian derived by Jackeli and Khaliullin \cite{PhysRevLett.102.017205} for layered iridates in the strong SOI limit. In fact, according to their theoretical model, when the tetragonal crystal field is not strong enough to modify  the $J_{\mathrm{eff}}=1/2$ state, magnetic and crystal structures are intimately related resulting in a perfect equivalence of the magnetic moment and octahedral rotation angles ($\phi=\rho$). This significant coupling between magnetic and structural degrees of freedom suggests the existence of a strong magnetoelastic effect, already  observed in \srsl \cite{PhysRevB.84.100402,PhysRevB.80.140407}. Using the error bar in the determination of the canting angle $\phi$ as an input parameter, together with the model Hamiltonian of Reference \cite{PhysRevLett.102.017205}, we can further extend  our analysis and set constraints on the effective tetragonal crystal field affecting the Ir$^{4+}$ ground state. Figure \ref{fig:CF} shows the tetragonal crystal field parameter $\Delta$ as a function of the SOI constant $\lambda$. For a typical value of $\lambda=0.42(5)$ eV in iridates \cite{PhysRevLett.110.117207,PhysRevLett.109.157401,PhysRevLett.110.076402}, we find $-60\,\mathrm{meV}\leq\Delta\leq 35\,\mathrm{meV}$, a value too small to induce a consistent deviation  from the pure $J_{\mathrm{eff}}=1/2$ picture in \srsl.

In conclusion, we have accurately determined the magnetic structure of \srsl using  X-ray resonant magnetic scattering. From a comparison between the observed intensity of several magnetic Bragg peaks  and the calculated resonant magnetic cross section for a model arrangement of Ir moments, we find that the Ir magnetic moments rigidly follow  the IrO$_6$ octahedra deviating by 12.2(8)$^{\circ}$ from the $a$-axis. Our results thus add to the growing weight of evidence   \cite{Kim-PhysRevLett-2008,Kim-Science-2009,PhysRevB.87.140406,PhysRevB.87.144405} that, in spite of the fact that the local environment of the Ir$^{4+}$ ions is distorted from perfect cubic symmetry, the ground state in \srsl has many of the attributes of the proposed $J_{\mathrm{eff}}=1/2$ state. More generally, the fact that the magnetic moment canting is slaved to the oxygen rotation, lends strong support to the  theoretical model which has been developed to understand the properties of this and other iridate perovskites \cite{PhysRevLett.102.017205}.

\section{Acknowledgments}
We would like to thank the Impact studentship programme, awarded jointly by UCL and Diamond Light Source for funding the thesis work of S. Boseggia. G. Nisbet provided excellent instrument support and advice on multiple scattering at the I16 beamline.  The work in the UK was supported through  grant EP/J016713/1 from the EPSRC, and in Denmark by the Nordea Fonden and the Otto M\o steds Fond. We also would like to thank F. Fabrizi and V. I. Schnells for helpful discussions.
\section{References}
\bibliographystyle{unsrt}
\bibliography{Sr214}
\end{document}